\documentclass{svjour3}                     
\smartqed  
\usepackage{graphicx}
 \usepackage{aps-bibstyle}  
\begin{document}

\title{Observational tests of the picture of disk accretion
}


\author{Thomas J. Maccarone}


\institute{T. Maccarone \at
              Department of Physics, Texas Tech University, Box 41051, Lubbock TX 79409-1051 \\
              Tel.: +1-806-742-3760\\
              Fax: +1-806-742-1182\\
              \email{thomas.maccarone@ttu.edu}           
}

\date{Received: date / Accepted: date}

\maketitle

\begin{abstract}
In this chapter, I present a summary of observational tests of the basic
picture of disk accretion.  An emphasis is placed on tests relevant to black
holes, but many of the fundamental results are drawn from studies of other
classes of systems.  Evidence is discussed for the basic structures of
accretion flows.  The cases of systems with and without accretion disks are
discussed, as is the evidence that disks actually form.  Also discussed are
the hot spots where accretion streams impact the disks, and the boundary
layers in the inner parts of systems where the accretors are not black holes.
The nature of slow, large amplitude variability is discussed.  It is shown
that some of the key predictions of the classical thermal-viscous ionization
instability model for producing outbursts are in excellent agreement with
observational results.  It is also show that there are systems whose outbursts
are extremely difficult to explain without invoking variations in the rate of
mass transfer from the donor star into the outer accretion disk, or tidally
induced variations in the mass transfer rates.  Finally, I briefly discuss
recent quasar microlensing measurements which give truly independent
constraints on the inner accretion geometry around black holes.
\end{abstract}

\section{Introduction}

Observations which, in hindsight, present evidence for mass transfer have been
known about since antiquity (e.g. the system Algol).  Evidence for accretion
power by black holes can be seen, again, in hindsight, as far back as the
discovery of jets from active galactic nuclei (Curtis 1918), and the bright,
broad-lined nuclei of the Seyfert galaxies (1943).  Theoretical discussion of
the possibility of accretion began in the 1940s (Kuiper 1941; Bondi \& Hoyle
1944), with the focus on mass transfer in contact binaries, and accretion from
the interstellar medium, respectively.  Studies of the importance of accretion
power in astrophysics first became prominent about 15 years later (Crawford \&
Kraft 1956), as it began to become clear that many of the classical, recurrent
and dwarf novae were short orbital period spectroscopic binaries, often with
photometric modulation at the same period (e.g. Walker 1954; Joy 1954; Joy
1956; Greenstein \& Kraft 1959; Kraft 1964).  The conclusion was thus drawn
that mass transfer must, in some way, be responsible for the unusual
phenomenology of the class of objects (see also Kopal 1959 for a synthetic
discussion).

The basic elements of accretion disk theory are reviewed in the chapter in
this volume by Omer Blaes, while a review of the progress in connecting the
``microphysics'' of magnetohydrodynamics to the ``macrophysics'' of classical
accretion disk theory is presented in the chapter by Chris Fragile.  This
article will focus on showing the observational tests that have verified that
simple pictures of how accretion disks should work, largely developed in the
1970s, match many of the observational constraints at some broad level.  It
will also discuss the observational evidence for cases where the simple
picture breaks down, and more complicated models must be invoked -- even
though in some cases it is not clear what those more complicated models are.

A variety of means of testing accretion disk theory can be made -- broadband
spectroscopic measurements (see e.g. Juri Poutanen's and Jeff McClintock's
articles in this volume), and measurements of rapid variablity (see e.g. the
article by Tomaso Belloni \& Luigi Stella in this volume) are generally used
to understand the processes in the inner accretion flows.  In some cases,
these methods also provide valuable information about the global accretion
process.  In this article, I will focus on discussing the observational
constraints on accretion disk theory that come from other methodology --
chiefly, but not exclusively variability on much longer timescales.  Apart
from some discussion of quasar microlensing, I will leave aside the discussion
of topics such as the detailed structure of inner accretion flows around black
holes, as this topic is covered elsewhere in this volume.  Accretion disk
theory will be discussed in this article in broadbrush strokes, to set the
stage for understanding which observations will be interesting, but will be
discussed in detail in the other articles in this volume.

\section{A basic picture: setting a target for observations to test}

In the standard picture of an accretion flow onto a compact object from a
binary companion, we have the following structures:
\begin{enumerate}
\item An accretion stream which begins at the donor star (see e.g. Albright \& Richards 1996 for a discussion of how evidence for streams can be found using Doppler tomography)

\item An accretion disk -- a geometrically thin, optically thick flow
  in Keplerian rotation, with a small inward drift due to either {\it
    bona fide} viscosity, or, as has become increasingly favored in
  recent years, a magnetic field effect that can be well modelled on a
  macroscopic scale as acting like viscosity.  The accretion disk is
  generally assumed to extend outward to the ``circularization
  radius'' -- the radius where the specific angular momentum of the
  accreted material is the same as that for gas in a circular orbit
  around the accretor.  If the stellar radius is larger than the
  circularization radius (as can happen for wind-fed systems, and for
  accretion by ``normal'' stars, rather than by compact objects), a
  disk will fail to form, and the accretion stream will impact the
  accretor directly.  A disk may also fail to form in the case of a
  rotating high magnetic field accretor, where the magnetic field of
  the central star becomes dynamically important for a circular disk
  outside the circularization radius (e.g. Ghosh \& Lamb 1979).

\item A ``hot spot'' where the accretion stream from the donor star
  collides with the outer accretion disk, and releases its excess
  kinetic energy.

\item In systems without black hole accretors, a boundary layer should
  exist, where the excess rotational energy of the innermost part of
  the accretion disk is dissipated (Lynden-Bell \& Pringle 1974).  In
  systems with black hole accretors, there may be excess rotational
  energy at the innermost stable circular orbit, but it should be
  transported across the event horizon.  As the focus of this volume
  is black holes, we will not discuss the boundary layers further,
  except to say that evidence for them is found in both white dwarf
  (e.g. Pandel et al. 2003) and neutron star (e.g. Mitsuda et
  al. 1984 \footnote{Many other authors decompose neutron star spectra
    in different ways -- White et al. 1986; Church \&
    Ba\/luci\'nska-Church 2004, and there remains debate about the
    right spectral models for accreting low magnetic field neutron
    stars.  There {\it is} widespread agreement that neutron star
    spectra usually require at least two thermal or quasi-thermal
    spectral components.}  accretors, but that their spectra can be
  complicated, and, often, the emission does not show up as an extra
  blackbody component added to a disk model fit (e.g. Godon \& Sion
  2005; Piraino et al. 1999).

\end{enumerate}

In active galactic nuclei, only the accretion disks are present.  Certain
types of deviations from this picture are well-studied, and represent separate
topics in their own right -- e.g. spectral state phenomenology (in particular
states in which the accretion disk becomes geometrically thick) and the
presence of jets.  These topics are covered in detail in other chapters in
this volume and will not be covered here in detail.

\section{Classes of accretors}
A wide variety of classes of accreting objects exist in the Universe.
While a large fraction of the literature on accretion involves the
study of mass-transferring binary stars, some accretion takes place
onto isolated objects.  In particular, active galactic nuclei are
generally presumed to accrete from their local interstellar medium.
In many protostars, the accreting object is a single star.  

Since the focus of this volume is on accretion onto black holes, the
focus of this article will be on the black hole accretors themselves.
Studies of stellar mass black holes are often plagued by small
number statistics.  Studies of active galactic nuclei suffer from
difficulty in making detailed measurements, as well as long timescales
of variability.  It is thus useful to supplement studies of black holes with
studies of other classes of accreting objects.  Additionally,
comparisons between black hole accretors and other classes of
accretors can be excellent ways to determine which phenomenology is truly
unique to black holes, rather than being generic to the process of
accretion.

The classes of mass transferring binaries seen in nature include:
\begin{enumerate}

\item Low Mass X-ray Binaries (LMXBs).  These are systems in which a
  neutron star or black hole accretes from a low mass main sequence
  star or a low mass subgiant star through Roche lobe overflow.  The mass
  transfer rates in these systems are determined by the rate at which either
  the donor star expands (due to nuclear evolution) or the orbit shrinks (due
  to magnetic braking and/or gravitational radiation).  
 
\item High Mass X-ray Binaries (HMXBs).  These are systems in which a neutron
  star or black hole accretes from a massive star.  Many of these objects have
  Be stars as the donor stars, and the accretion seems to take place from the
  equatorial wind.  Others have supergiant donors, and the accretion takes
  place through gravitational capture of the stellar wind of the supergiant.
  For these fast-wind systems, the accretion disk may begin to have a circular
  orbit very close to the compact object, or, in some cases, a disk may not
  form at all.  Finally, others have much shorter orbital periods, and their
  donor stars are either Roche-lobe filling or nearly Roche-lobe filling.  In
  these systems the mass transfer takes place through either a focused wind,
  or actual Roche lobe overflow.  See Corbet (1986) for a discussion of the
  different classes of high mass X-ray binaries.

\item Symbiotic stars.  These are systems which, based on their initial
  definition, show evidence for both hot and cool components in their optical
  spectra.  Since their initial discovery, it has been realized that the cool
  components are red giants, and the hot components are various forms of
  accreting objects.  Most symbiotic stars have white dwarf accretors, but a
  small fraction have neutron star accretors (Hynes et al. 2013 and references
  within).  The mass transfer is generally believed to take place through
  capture of the red giant star wind, although some symbiotic stars are at
  least very close to being in Roche lobe contact.  For a relatively recent
  review, see Sokoloski (2003).

\item Cataclysmic variables.  These are systems in which mass is
  transferred from a low mass main sequence star or a low mass
  subgiant star to a white dwarf; they are the analogs of low mass
  X-ray binaries, for systems where the accretors is a white dwarf.
  Cataclysmic variables are broken into a large number of subclasses
  based on different observed phenomenology of variability.  These
  sub-types are often named in terms of the prototype object, as is
  typical for nomenclature of variable stars.  In this article, we
  will use nomenclature descriptive of the phenomenology, a practice
  which is thankfully becoming more common in the cataclysmic variable
  community as well.

\item Ultracompact binaries.  These are binaries in which the donor
  star is degenerate -- either a white dwarf, or a low mass degenerate
  helium star.  In the Milky Way, these have been seen only with white
  dwarf or neutron star companions.  When the system is a double white
  dwarf binary with mass transfer, it is called an AM CVn star, after
  the prototype object.  When the accretor is a neutron star (or,
  potentially, a black hole) it is called an ultracompact X-ray
  binary.  We refer the reader to Nelemans \& Jonker (2010) for a
  review on the ultracompact X-ray binaries, and to Maccarone et
  al. (2007) and Zepf et al. (2008) for a discussion of the evidence
  for an ultracompact black hole X-ray binary in NGC~4472.

\item Various classes of close binaries with two ``normal'' (i.e. not
  compact remnant) stars show evidence for accretion.  These include,
  for example, Algol systems (in which mass transfer takes place from
  an evolved star to a main sequence star) and W~UMa systems (contact
  binaries, in which both stars overflow their Roche lobes
  simultaneously).  We refer the reader to Thomas (1977) for a review
  of these systems. Accretion disks form rarely in these systems
  (although see Olson 1991 for evidence of a disk in one Algol,
  KU~Cyg).

\end{enumerate}

While active galactic nuclei represent one of the two major classes of
accreting black holes, almost no fundamental tests of accretion theory
have been put forth primarily from studies of active galactic nuclei.
Such tests would be exceedingly difficult -- tests based on
spectroscopy would run into the problem that these systems are not
fully ionized like the disks of X-ray binaries, making models of the
inner accretion disks much more complicated (e.g. Done et al. 2012) --
trying to understand the inner accretion disks around black holes
using active galactic nuclei is considerably more difficult than
trying to understand the inner disks around stellar mass black holes.

A further large part of our understanding of the outer parts of accretion
disks comes from studies of variability on timescales long compared with the
light crossing time at the Schwarzschild radius.  In this case, the problem
for using active galactic nuclei as test cases stems from the fact that their
viscous timescales, even at the inner edge of the accretion disk, are expected
to be much longer than the durations of any light curves assembled by
astronomers.  As a result, nearly all of our understanding of the outer parts
of accretion disks, as well, comes from studies of stellar mass accretors.  It
should be noted, of course, that many fundamental advances in the studies of
outflows from accretion disks have been developed with essential contributions
from observations of active galactic nuclei, and in recent years, studies of
lensed quasars have started to give some distinctive observational tests of
the accretion geometry in active galactic nuclei.

It is thus the case that most of the fundamental constraints on
accretion disk theory must come from studies of mass transferring
binary systems, because they present accessible timescales, and have
disks in an ionization state which is simpler than do active galactic
nuclei.  Our goals should be to develop fundamental theories of
accretion in general, and then to determine which phenomenology is
specific to black holes.  As a result, it often makes sense to
incorporate observational constraints from other classes of accreting
objects.  The systems with neutron star accretors are the most similar
to those with black hole accretors, given that the radiative
efficiency for accretion onto a neutron star is very similar to that
for a non-rotating black hole, at least in the context of a standard
Shakura-Sunyaev accretion flow.

The systems with white dwarf accretors (and particularly the cataclysmic
variable stars, rather than the symbiotic binaries) are, however, the class of
systems which have often provided the best constraints on accretion theory.
Like black hole and neutron star accretors, the cataclysmic variables have the
emission from their primary stars dominated by the accretion
process,\footnote{One exception is the surface layer fusion that can take
  place in supersoft sources.  Another exception is in classical nova
  explosions.  Novae can actually dominate the total energy output from
  accreting white dwarfs, but they have very low duty cycles, and hence are
  negligible most of the time for most CVs.}  rather than by core fusion, so
that the radiation from the accretor can be taken as a tracer of the accretion
disk's activity.  That nuclear fusion contributes significantly to the
emission only in low-duty cycle bursts is a fundamental difference between
accreting compact objects and other kinds of accreting stars.

It may seem a bit odd for a volume on black holes to include a substantial
disucssion of the literature on cataclysmic variables, but in many cases, the
CVs provide tighter constraints on the basic physics processes which can be
expected to be generalizable to all of accretion physics.  The chief advantage
of cataclysmic variables relative to X-ray binaries for studying accretion is
that the CVs are more numerous.  About 10 times as many CVs as LMXBs are
known, and the nearest CVs are about 10 times as close as the nearest low mass
X-ray binaries.  While in recent years, there have been a few geometric
parallax measurements made of X-ray binaries in the radio (e.g. Miller-Jones
et al. 2009; Reid et al. 2011), such measurements have been made for much
larger samples of CVs.  It also turns out that the outbursting cataclysmic
variable stars tend to have shorter recurrence times than the outbursts X-ray
binaries, allowing for a large class of systems which have been studied
repeatedly.  The other advantages of the CVs being closer is that they are
brighter optically, and are observed at lower extinction.  Furthermore by
being more numerous, there are many more eclipsing CVs known than eclipsing
neutron star, or, especially, black hole X-ray binaries.

\section{Evidence for the basic structures of accretion disks}

A variety of observational approaches has been used to establish the basic
picture of disk accretion seen in binary systems.  These include both
spectroscopic observations, and variability studies.  Here, we first show that
accretion disks really do form in many cases, and discuss the circumstances
where accretion takes place without disks.  We then discuss techniques that
can be used to map out the accretion geometry in binary systems.

\subsection{Proof that accretion really happens in disks}
Among the first things worth testing are whether accretion really does
take place in disks.  In fact, the paper with the first association of
stellar binarity with the production of classical novae (Walker 1954)
presents a key piece of evidence that the accretion process, at least
in those systems must take place via a disk.  It shows the properties
of the eclipse of the system -- relatively long ingresses and egresses
and short minima for the primary eclipse, while having much weaker
secondary eclipses (constrained by Walker \& Herbig 1956 to be less
than 0.03 magnitudes, with primary eclipses of about 1 magnitude).  In
hindsight, these eclipse properties clearly indicate that the solid
angle subtended by the accretion disk must be a small fraction of what
would be subtended by a star of the same maximum radial extent.  A
second, more direct, but more recent piece of strong evidence that the
objects believed to be accretion disks really are disks is that many
of them show double-peaked emission lines (see Bailey \& Ward 1981;
Marsh 1988 for a discussion of cataclysmic variables; Eracleous \&
Halpern 2003 for a discussion of active galactic nuclei; and Soria
2002 for a discussion of X-ray binaries).  

\subsubsection{Systems with accretion not happening via disks}
At the same time, there are clear examples of systems in which there
is accretion which {\it does not} take place through a disk.  These
are systems which have circularization radii smaller than the radius
at which a potential accretion disk would be disrupted.  The most
obvious such radius would be the radius of the accreting star, and
indeed for non-compact stars, accretion disks are the exception rather
than the rule.  For compact stars, the relevant radius will usually be
the magnetospheric radius of the accretor.

The observational data on accreting objects support this theoretical
scenario.  There are classes of accreting white dwarfs and accreting
neutron stars which lack the standard signatures of disk accretion.
The accreting white dwarfs in this class are called polars.  The name
derives from the fact that their accretion light is often strongly
polarized and that polarization stems from the fact that the accretion
is channelled down the systems' magnetic poles.  These systems release
large fractions of their emission in the X-rays relative to other
cataclysmic variables, because the emission comes mostly from an
accretion column rather than accretion disk.  They also frequently
show periodic emission, with the modulation taking place on the
rotation period of the white dwarf.  An analogous class of neutron
stars are the accretion-powered X-ray pulsars.  In both classes of
objects, cyclotron lines have been seen (e.g. Reimers \& Hagen 2000
for polars; Hemphill et al. 2013 for X-ray pulsars).  It is important
to note that there are classes of systems which show magnetically
dominated accretion and disk accretion at the same time -- the
intermediate polars among cataclysmic variables (Warner 1983), and
both slow (Jonker \& van der Klis 2001; La Barbera et al. 2001) and
millisecond (Wijnands \& van der Klis 1998) X-ray pulsars among the
neutron stars.

\subsection{Eclipse mapping of accretion disks}

At this point we have established that disks really do form in
accreting objects.  We also have sound theoretical reasoning, combined
with empirical support, to show that the binary systems which do not
have accretion disks have accretors which are fundamentally different
from black holes -- they either have surfaces at large radii, or they
have dynamically important magnetic fields.  It is thus reasonable to
assume that all accretion onto black holes takes place through
accretion disks, and, in this volume about black holes, to worry only
about disk accretion from this point on.

Now, we can determine whether certain specific predictions of the simplest
accretion disk models are in agreement with the observational data.  One of
the most straightforward predictions is that any disk in a steady state should
have $T\propto{R^{-3/4}}$. This result comes from equating the differential
blackbody luminosity in an annulus, $dL = 2\pi{\sigma}T^4 R dR$ to the
differential power released by gas falling through that annulus,
$\frac{GM\dot{m}}{R^2}dR$, and solving for $T$ as a function of $R$.  In this
framework $\dot{m}$, the accretion rate, and $M$, the accretor mass, are
constants in a steady state accretion disk, while $T$ is the temperature of
the annulus, $R$ is the radial distance of the annulus from the compact
object's center, $dL$ is the luminosity of the annulus, and $G$ and $\sigma$
are the gravitational constant and the Stefan-Boltzmann constant,
respectively.

The technique of choice for testing these models has been eclipse mapping of
cataclysmic variables (e.g. Horne 1993; Baptista 2001).  Eclipse mapping is
one of the few means of getting geometric information about continuum emission
processes of accretion disks (with quasar microlensing being the other major
method).  Cataclysmic variables are the system class of choice for this work
because there are optically bright eclipsing cataclysmic variables.  No
eclipsing low mass X-ray binaries with black holes are known, and the few
eclipsing neutron star X-ray binaries are not as bright as the brightest
eclipsing CVs.  Furthermore, bright accreting neutron stars are likely to have
important effects from irradiation of the outer accretion disk by the inner
accretion disk, meaning that the implications of a disagreement between the
data and the standard theoretical $T\propto{R^{-3/4}}$ law might be expected,
and could be difficult to disentangle from other effects.

It is most convenient to begin by attempting to model the ``novalike''
(i.e. persistently bright) cataclysmic variables, or the ``dwarf
nova'' cataclysmic variables near the peaks of their outbursts.  These
are the systems that are expected to have nearly constant mass
transfer rates on timescales of order the viscous propagation
timescale through the accretion disks.  The novalike systems have
frequently failed to show $R^{-3/4}$ temperature profiles (Wood et
al. 1992a; Baptista et al. 1995 -- but see Rutten et al. 1992 for an
alternative result); they often typically show much flatter
temperature profiles. The quiescent dwarf novae almost universally
show temperature profiles that are much flatter than $R^{-3/4}$
(e.g. Wood et al. 1986,1992a).  This can be interpreted as a build-up
of mass in the outer accretion disk during quiescence, a loss of mass
due to winds which take mass away from the inner disk relative to what
is in the outer disk, the emission of that optically thin disk wind
which emits substantial light, or some combination of the different
effects (e.g. Wood et al. 1986; Baptista et al. 1998).

Knigge et al. (1998) showed that the integrated spectra found simply
from summing optically thick blackbodies does not describe the
integrated spectrum of UX UMa, one of the prototype objects for
eclipse mapping studies.  Baptista et al. (1998) and Robinson et
al. (1999) show that the results of eclipse mapping campaigns can be
affected significantly by errors in what had previously been standard
treatments -- the treatment of brightness temperatures as effective
temperatures (implictly assuming optically thick blackbody emission as
the only source of light), and the failure to compute the effects of
limb darkening properly.  They find that even with careful treatment
of limb darkening, the temperature profiles are flatter than
$R^{-3/4}$.  

Additionally, with a more careful treatment of the radiative transfer
in accretion disks, it becomes possible to model the vertical
structure of the disks.  The standard Shakura-Sunyaev treatment yields
a ratio of height $H$ to radius $R$ of $H/R \propto R^{1/8}$.
Cataclysmic variable accretion disks typically span a range of a
factor of only about 10-100 in radius between the surface of the white
dwarf and the outer edge of the accretion disk, meaning that $H/R$
should change by, at most, a factor of 1.8.  

Only relatively large values of $H/R$ can be measured using eclipse
mapping techniques, so generally, attempts are made only to estimate
the scale height of the outer accretion disk.  The numerical values of
the theoretical scale heights for the Shakura-Sunyaev model indicate
that $H/R\sim 0.03$ should be typical for bright CVs.  Higher values
have generally been found (e.g. $H/R$ of about 0.06 for Z Cha),
indicating that some additional process is puffing up the outer disks
in these systems, or that some other geometric feature or radiative
transfer process is not accounted for in the existing eclipse mapping
analysis (Robinson et al. 1999).

A few eclipse mapping studies of X-ray binary accretion disks have been made
as well.  Here, one expects irrations of the outer disk by the inner X-ray
emitter to heat the outer disk, and cause it to have a larger scale height
than expected in the context of the Shakura-Sunyaev disk model (e.g. Meyer \&
Meyer-Hofmeister 1982).  At least for the source X~1822-371, the prediction of
Meyer \& Meyer-Hofmeister (1982) is verified (Puchnariewicz et al. 1995;
Bayless et al. 2010).  X-ray eclipse mapping has shown large spatial scale
X-ray emission.  This is sometimes interpreted in terms of the region in which
the X-rays are produced being spatially very large (Church 2004), but may be
due to large scale optically thin disk winds which scatter a small fraction of
the X-ray emission back into the observer's line of sight.  It is generally
true in X-ray binaries that the disk winds seem to be more important in soft
states than in hard states (Neilsen \& Lee 2009); it is also true that the
accretion disk corona sizes from eclipse mapping are larger in bright sources
than in fainter sources (Church 2004).

To date, a single strong candidate eclipsing black hole X-ray binary
is known (see Pietsch et al. 2006 for evidence of the eclipsing nature
of the object and Orosz et al. 2007 for the dynamical evidence that
the object is a black hole X-ray binary).  This system is a high mass
X-ray binary with a luminous 70 $M_\odot$ donor star, (Orosz et
al. 2007).  The combination of the brightness of the donor star
relative to the accretion disk, and the fact that the donor star
should have a strong wind, and hence not act as a ``sharp edge'' for
doing eclipsing mean that eclipse mapping of this accretion disk is
not particularly promising.  Furthermore, the object is in M33, at a
distance of about 800 kpc.

\subsection{Evidence for hot spots}
There are multiple lines of reasoning supporting the existence of hot spots
where accretion streams impact the outer circular disks of accreting objects.
In general, the hot spots in X-ray binaries can be quite a bit more difficult
to detect than those in cataclysmic variables.  This can be well understood in
terms of the fraction of the total energy released as the material falls
inwards.  If one sets the expected luminosity of an accretion flow due to a
fall through a potential from height $r_{out}$ to height $r_{in}$, then
$L=-GM\dot{m}\left(\frac{1}{r_{out}}-\frac{1}{r_{in}}\right)$.  The hot spot
luminosity can be obtained by setting $r_{out}$ to the orbital radius, and
$r_{in}$ to the circularization radius, while the total luminosity can be
obtained by setting $r_{in}$ to the radius of the compact star.  For systems
with orbital periods of a few hours, the few $\times10^9$ cm radii of white
dwarfs will typically be of order 10\% of the circularization radii, so a
substantial fraction of the luminosity will be produced at the hot spot.  For
X-ray binaries, the fraction of the power produced at the hot spot will be a
factor of order 1000 smaller.  Hot spots can thus be detected in black hole
and neutron star accretors only if either the mass transfer rate is extremely
low, and hence the radiative efficiency of the inner accretion flow is
extremely small, or in the more common case, the systems are transients, and
are being observed during a quiescent period, in which the accretion rate into
the outer accretion disk far exceeds the accretion rate onto the central
compact object (McClintock et al. 1995; Froning et al. 2011).  The hot spots
tend to have a larger vertical height from the disk midplane than do the other
parts of the outer disk, so there are also cases, for specific inclination
angles, where the hot spot occults the inner disk when it is in the observers
path to the compact object (White \& Mason 1985).  In quiescent dwarf novae,
the hot spot luminosity can be a very large fraction of the total luminosity
from the system, leading to strong orbital modulations as the viewing angle of
the hot spot changes (e.g. Wood et al. 1989).

\subsection{Spiral structure: evidence for deviations from simple disk models}

The technique of Doppler tomography (Marsh \& Horne 1988) allows a
form of indirect imaging of accretion disks by looking at how line
profiles change as a function of orbital phase.  Steeghs (2000) shows
two-armed asymmetries in a several cataclysmic variables in outburst
using Doppler tomography.  The evidence for such phenomena in X-ray
binaries is much weaker, although it has been suggested that spiral
density wave may help explain some of the large amplitude variability
in GRS~1915+105 (Tagger \& Pellat 1999).

\section{Large amplitude, long timescale variability}

One of the most important facets of the behavior of both cataclysmic
variables and X-ray binaries is the presents of large amplitude,
relatively smooth variations.  These are often called dwarf nova
outbursts in cataclysmic variables, and X-ray novae, or soft X-ray
transient outbursts, in the X-ray binaries.  In the X-ray binaries,
these transients can lead to variations in the X-ray luminosities of
the accreting systems of factors of $10^4 - 10^6$ or more.  

As it became clear that the dwarf novae and classical novae, often
seen in the same objects, were fundamentally different
phenomena,\footnote{Classical novae are runaway nuclear fusion
  episodes on the surface of white dwarfs (Schatzman 1949), and hence
  have nothing to do with accretion disks, apart from that disks are
  usually the means by which the gas is transported to the white
  dwarf.} two models emerged for explaining the dwarf novae.  Both
involved modulating the accretion rate onto the compact object.  In
one model, the mass transfer instability model, the rate at which mass
{\it enters} the accretion disk is variable, while in the other model,
the disk instability model, mass is supplied to the accretion disk at
a constant rate, but there are instabilities in the way the mass flows
through the disk.  I will argue in this article that there is evidence
supporting, if not demonstrating conclusively, the idea that both of
these mechanisms apply at least some of the time.

\subsection{Mechanisms for large variations in luminosity}
Some mechanisms for producing mass transfer rate variations may be
irradiation of the donor star by the accretor (e.g. Hameury et
al. 1986; Harpaz \& Rappaport 1991), magnetic activity cycles in the
donor stars (e.g. Bianchini 1990), or changes in the eccentricity of
the orbit of the inner mass transferring binary in a hierarchical
triple system (e.g. Hut \& Paczynski 1984; Maccarone 2005; Zdziarski
et al. 2007).  Mass transfer rates depend on the gas density at the
inner Lagrange point.  The pressure scale height in a stellar
atmosphere is typically of order $10^{-4}$ of the orbital separation.
Therefore, changes in either the radius of the star or the orbital
separation of the binary can lead to large changes in the mass
accretion rate of order $10^{-4}$ could lead to factors of a few
changes in the mass transfer rate.  Thus, the lack of direct evidence
that these changes should not be taken as proof that the mass transfer
rate is not changing via these mechanisms.  Strong evidence for mass
transfer variations can come in the form of finding variations in the
accretion rate when one averages over timescales much longer than any
reasonable timescale for mass to propagate through the accretion disk.
This does not preclude variations in the mass transfer rate which
happen faster than this timescale -- it is just that such faster
variations are likely to be extremely difficult to disentangle from
disk instabilities, and as we will show in this article, considerable
evidence exists that disk instabilities do explain much of the
outburst phenomenology in accretion disks.

The thermal viscous instability is strongly favored as the primary source of
disk instability.  In this scenario, the viscosity parameter of the accretion
disk is a function of the ionization state of the gas, so that when the disk
is cold and neutral, the mass flow rate is smaller than when the disk is hot
and ionized.  Given, also, a temperature-density relationship for the disk, a
limit cycle instability will develop for accretion rates within a range
commonly seen in binaries with accreting compact objects, so that many of
these objects are expected to show outburst cycles as predicted by the limit
cycles.  The leading alternative, or perhaps complement, to the disk
instability model is a model in which the mass transfer rate is varied.  An
extensive review of the features, successes, and shortcomings of the disk
instability model is given by Lasota (2001).  I will summarize some of the
major points presented in that work, but will focus in this section on the
observational developments since that time.

\subsection{The thermal-viscous instability and stability criteria}
The basic first order predictions of the disk instability model are seen to be
followed pretty well by most transient accreting compact objects.  By and
large, the systems which have accretion rates high enough that one would
expect their disks always to be fully ionized are persistent, and the systems
which have accretion rates lower than that value are transient (Lasota 2001;
see Coriat et al. 2012 for a confirmation that the result still holds with a
much larger sample of objects). A particular system near the threshold for
stability is persistently accreting, but shows large amplitude variability
(Maccarone et al. 2010).

An encouraging recent result came from the measurement of a precise geometric
parallax distance for SS~Cyg.  For quite some time it had simultaneously been
held up as the prototype system for studying dwarf nova outbursts, and had
been a system which appeared to have too high a mass accretion rate to allow
the dwarf nova ionization instability to take place.  The VLBI parallax
distance found by Miller-Jones et al. (2013) indicates that the source is
closer than was previously thought.  The change of distance results in a mass
transfer rate for SS~Cyg sufficiently low that the system is below the
threshold for stable accretion in the ionization instability model.

\subsection{Disk instabilities and peak outburst luminosities}
Warner (1987) first found the relation between peak brightness and
orbital period for cataclysmic variables, and that has recently been
revisited with a larger sample by Patterson (2011) who finds:
\begin{equation}
M_{V,peak} = 5.70 - 0.287 P_{orb,hr}.
\end{equation}
A least-squares fit of the data from Patterson (2011) to a power law
relationship finds that the peak luminosity scales with
$P_{orb}^{1.2}$.  These data are plotted in Figure 1.  Given the
relatively large scatter in the data, it is in fairly good agreement
with the predictions for theoretical models which suggest that the
whole accretion disk should be at a constant temperature in dwarf
novae outbursts (e.g. Osaki 1996; Cannizzo 1998; Smak 2000), yielding
a $L\propto{P}^{4/3}$ relationship.

Some other indications that favor disk instabilities as a baseline
model come from looking at the peak luminosities seen from X-ray
binaries and from cataclysmic variables.  In both cases, these are
well-correlated with orbital period (Shahbaz et al. 1998; Portegies
Zwart et al. 2004 -- P04; Wu et al. 2010).  Wu et al. find:
\begin{equation}
\frac{L_{peak}}{L_{Edd}} = -1.80 + 0.64 {\rm log} P_{orb,days}
\end{equation}
for X-ray binaries, althought with a different treatment of the bolometric
corrections, P04 found a steeper relationship for short orbital periods and a
saturation at about the Eddington luminosity for long orbital periods.  These
data are plotted in Figure 2.  In any event, there is at least rough agreement
with the finding of King \& Ritter (1998) that outburst peak luminosities
should scale with the radius of the accretion disk.

\begin{figure}
  \includegraphics[angle=-90,width=12 cm]{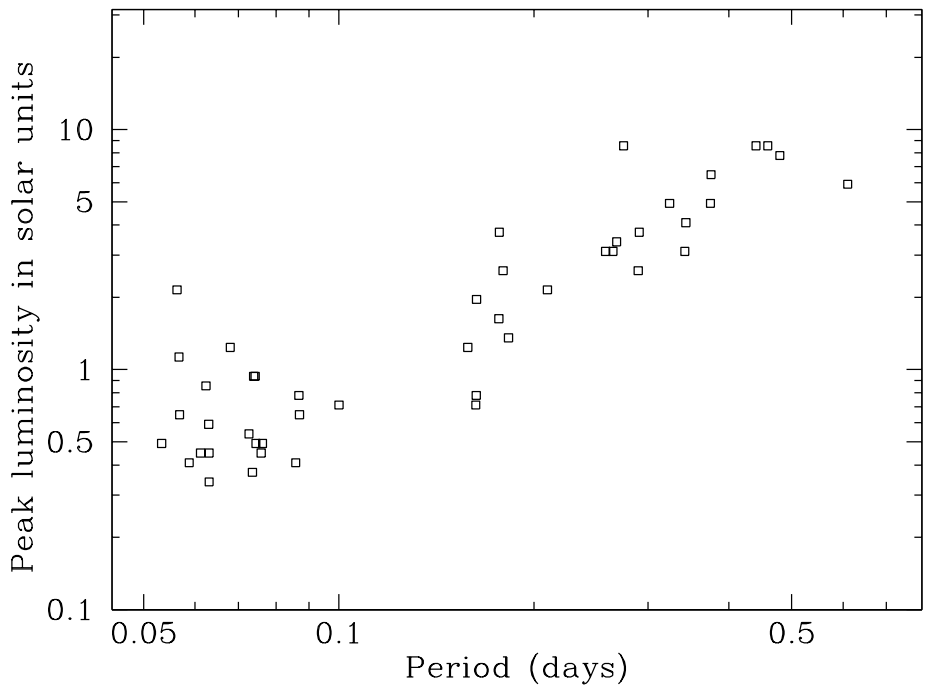}
\caption{The relation between peak optical luminosity and orbital
  period for dwarf novae.  The data are taken from Patterson (2011).
  The relation is consistent with the $L\propto{P}^{4/3}$ relation
  expected from theory.  Data points are plotted without error bars
  for clarity, but the typical errors are $\sim$10-20\% on the peak
  luminosity, and negligible on the periods.}
\label{pattersonfigure}       
\end{figure}

\begin{figure}
  \includegraphics[angle=-90,width=12 cm]{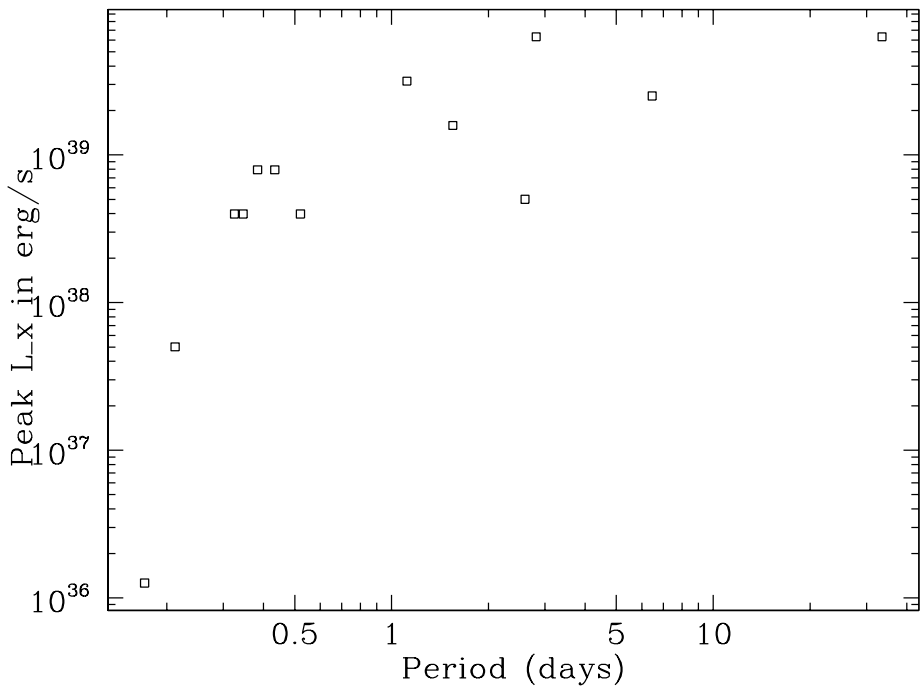}
\caption{A replotting of the data presented in Portegies Zwart et
  al. (2004), which shows that the peak X-ray luminosity for X-ray
  transients increases with orbital period.  Unlike for the
  cataclysmic variables, irradiation is expected to be important, the
  distances are in many cases very poorly known, the bolometric
  corrections can be uncertain, and some systems may remain in
  radiatively inefficient states, so it is difficult to make
  quantitative comparisons between theory and data.}
\label{pzdmfigure}
\end{figure}

\subsection{Outburst durations}
The durations of outbursts of many, but not all, systems, are relatively well
in-line with the expected viscous timescales of the accretion disks for the
black hole systems (see e.g. Chen, Shrader \& Livio 1997; P04).  The outburst
durations are shorter than the viscous timescales for the CV accretion disk,
where irradiation is not important, and so cooling fronts can truncate the
outbursts before the entire disks are accreted (Lasota 2001).  Some
significant amount of the data which represent exceptions to the basic
ionization instability may be explained as the result of tidal effects (see
section 5.5.1).

\subsection{Phenomena which are hard to explain in terms of disk ionization instabilities}
While the ionization instability explains the phenomenology of X-ray binary
and CV outbursts in broad brushstrokes, there are phenomena which are clearly
not in agreement with that picture.  In the cataclysmic variables, where the
recurrence times between outbursts tend to be much shorter than in X-ray
binaries, it can be clearly seen that there are variations from outburst to
outburst in ways that have been fit, to date, only by adding in truncations of
the inner accretion disk {\it and} variations of the mass transfer rate
(e.g. Schreiber et al. 2003; Lasota 2012).

There are a few other cases where strong evidence for mass transfer
instabilities are expected.  A prime recent example among black hole
candidates is the ongoing outburst of Swift J1753.5-0127, which has been in
outburst since 2005, and has an orbital period of about 3.2 hours (Zurita et
al. 2008).  The outburst duration of 8 years (and counting) combined with the
short orbital period is something that cannot be explained in terms of an
ionization instability model.  While the observation of superhumps (see the
following section for a discussion of superhumps) in this source (Zurita et
al. 2008) should imply that this outburst is a ``super-outburst'' and hence
should be longer than normal outbursts, the super-outbursts in well-studied
systems are only a factor of a few longer than the normal outbursts.  Several
neutron star accretors have also undergone outbursts that lasted far longer
than the expected viscous timescales for the systems' orbital periods
(e.g. Wijnands et al. 2001 and references within).

On the flip side, the 1999 transient episode of the accreting black hole XTE
J1819-254 showed strong evolution on timescales of a few hours, having reached
a flux of about 12 Crab, corresponding to a super-Eddington luminosity for the
source (Hjellming et al. 2000; Orosz et al. 2001).  The variations were
clearly too fast to be the result of some global disk instability, and too
strong to have been the result of the normal variability typically seen in
X-ray binaries.

An additional line of evidence for variations in the mass transfer rates --
perhaps the most direct such evidence -- comes from Cantrell et al. (2010).
They interpret the blue excess in the light from A~0620-00 in quiescence as
coming from the accretion impact spot.  In the context of that interpretation,
the variability in the quiescent luminosity of the hot spot immediately
implies that the rate at which matter is reaching the outer accretion disk is
changing strongly as a function of time.  The variation in the quiescent
ultraviolet flux of the source also supports this interpretation, since the
ultraviolet light can be demonstrated even more convincingly than the optical
light to come from the hot spot (McClintock et al. 1995; Froning et al. 2011).

\begin{figure}
  \includegraphics[angle=-90,width=12 cm]{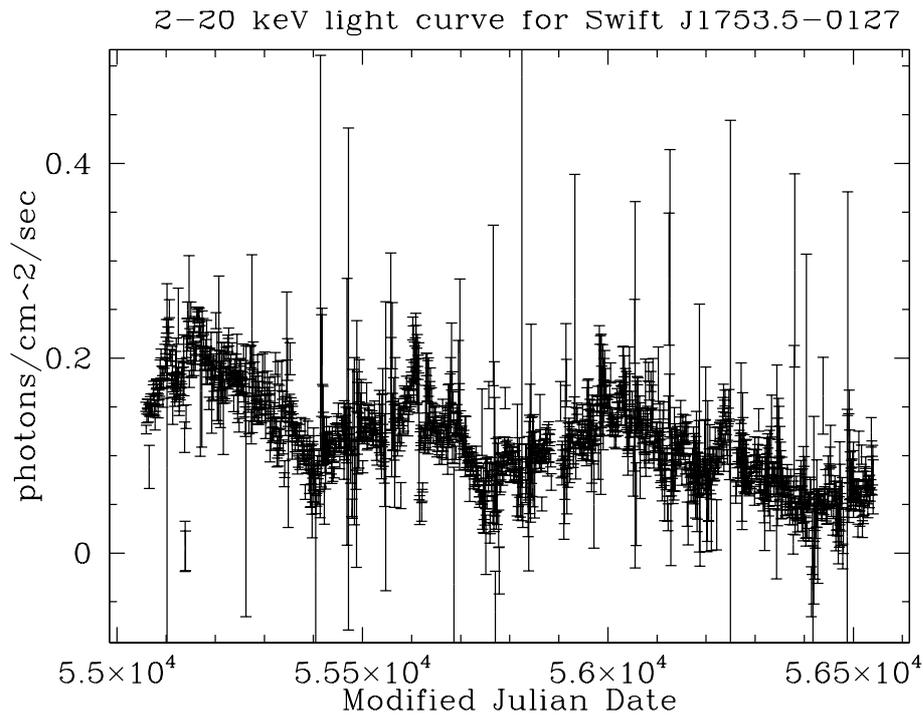}
\caption{The MAXI data for Swift J1753.5-0127.  The system, which had
  not been observed prior to 2005, has clearly been a bright X-ray
  source for the duration of the MAXI mission.  Given its orbital
  period of 3.2 hours, its viscous timescale should be much less than
  a year, and the long outburst cannot be explained in the context of
  the standard disk instability model.}
\label{maxifigure}       
\end{figure}

\subsubsection{Tidal effects}

In an X-ray binary, the presence of a donar star means that tidal forces on
the accretion disk may be substantial and variable on the orbital period.
There are a few observe phenomena which have very well motivated theoretical
explanations as coming from tidal effects.  There are also a few pheneomena
which may more speculatively be associated with tidal effects.

One phenomenon which is well-associated with tidal interactions is that of
superhumping.  Superhumps are oscillations in the light curves of some
outbursting dwarf novae and X-ray binaries.  There is a critical mass ratio of
0.35 below which CVs in bright states show superhumps, and above which CVs
never show superhumps (Patterson et al. 2005).  The oscillations occur with a
period very close to, but not exactly equal to, the orbital period of the
binary.  The oscillations are well-explain by a model in which an eccentric
instability develops at the 3:1 resonance between the orbital period of the
accretion disk and the orbital period of the binary system, and then this disk
precesses due to the tidal forces (Whitehurst 1988).  The limit on the mass
ratio is then given by the limit that the accretion disk not be tidally
truncated inside the location of the 3:1 resonance region.

The superhumping behavior is associated with ``super-outbursts'' of these
systems.  Most of the cataclysmic variables which show superhumps show a
sequence of outbursts, in which some small fraction of the outbursts are
significantly brighter than the rest of them.  For example, in V1504 Cyg,
which has been well-studied with Kepler, the superoutbursts happen about once
for every 14 normal outbursts (Osaki \& Kato 2013).  

The prevailing view for these phenomena being so well coupled is that the
thermal-tidal disk instability explains both effects (e.g. Osaki 1996).  In
this model, the outbursts of the disk are caused by the standard thermal
instability model, discussed above.  The outer edge of the accretion disk
moves outwards during each normal outburst.  After a series of normal
outbursts, the disk is outside the 3:1 resonance radius, and the next normal
outburst triggers the tidal instability, which drives in material from far out
in the disk, leading to an increased peak accretion rate and outburst
duration.  A modification of the tidal instability model has been proposed by
Truss et al. (2002) to explain the re-brightening in soft X-ray transients.
In their model, they suggest that irradiation of the outer accretion disk is
uneven.  Tidal forces cause the location of the irradiation region to change,
allowing for the originally un-irradiated region to become irradiated later,
allowing, at a relatively late time, a new portion of the disk to become hot
and enter a high viscosity state.

Superhumps have also been seen in several short period X-ray binaries
(e.g. O'Donoghue \& Charles 1996; Zurita et al. 2002; Zurita et al. 2008; Wang
\& Chakrabarty 2010; see also Wachter et al. 2002 for a more tentative
result).  The emission modulation mechanism in the CVs is thought to be tidal
modulation of the viscosity, a mechanism which cannot work in low mass X-ray
binaries because such a small fraction of the energy is dissipated in the
outer accretion disks of X-ray binaries (e.g. Haswell et al. 2001).

Among the X-ray binaries which have shown some evidence for superhumps are
short period black hole X-ray binaries (e.g. Nova Mus 1991, GRO J0422+32,
XTE~J1118+480 and Swift~J1753.5-0127), and a few recurrent transient neutron
star X-ray binaries (4U~1608-52 -- Wachter et al., 2002; and Aql X-1 --
R. Jain, private communication -- and see also the discussion in Wachter et
al. 2002 ).  Interestingly, the two recurrent transient neutron stars are
known to show outbursts with different amplitudes -- e.g. Aql X-1 seems to
show some kind of outburst roughly every 100 days, and a more major outburst
every 300 or so days (Maitra \& Bailyn 2008).  There is also evidence for
superhumping in the ultracompact X-ray binary 4U~1820-30 (Wang \& Chakrabarty
2010) -- this system is {\it persistently} bright, and at a short orbital
period.

The black holes which have been seen to show prominent superhumps in recent
years have not shown outbursts of varied amplitude.  It may be, however, that
these systems have shown {\it only} super-outbursts.  This idea has been put
forth by Maccarone \& Patruno (2013) as part of the reason why some short period
black hole X-ray binaries seem to be brighter than one might expect given the
Wu et al. (2010) relation, and might then allow for the normal outbursts of the
short period black hole X-ray binaries to manifest themselves as very faint
X-ray transients (see also Knevitt et al. 2013 who discuss how short period
transients may be absent from all-sky surveys).  Such a scenario is strongly
bolstered by finding superhumps in the bright outbursts from short-period
systems.  It is worth further noting that the ``normal'' outbursts from such
systems, because of radiative inefficiency for black holes in the low hard
state, should be $\sim10$ times fainter than their superoutbursts, rather than
just the factor of a few difference seen for the CV superoutbursts and the
candidate superoutbursts in the neutron star X-ray binaries.

\subsection{Large amplitude variability in active galactic nuclei}
A much harder question to answer is whether active galactic nuclei are
susceptible to the same type of accretion disk instabilities as X-ray
binaries.  In principle, they should be, as their accretion disks are much
cooler than those of X-ray binaries, but theoretical calculations suggest that
the outbursts may be quite a bit less dramatic than in X-ray binaries (Hameury
et al. 2009).  Determining observationally whether they do have outbursts is
complicated by the fact that the typical one month timescales for outbursts of
X-ray binaries, if scaled up to even the smallest, $10^6 M_\odot$ black holes
in AGN, would take place on timescales of several millenia.  Evidence for a
rather sharp variation in the luminosities of the Galactic Center can be seen
by looking at reflection spectra from molecular clouds, which indicate that it
was several orders of magnitude brighter about 100 years ago than it is now
(e.g. Ponti et al. 2010) -- still, there is no observational means to
determine whether this variation was due to changes in the mass transfer rate
into the AGN's accretion disk, or changes in the rate of flow through the
disk.  Koerding et al. (2006) found that the spectra of active galactic nuclei
are consistent with following a hysteresis loop like that followed for black
hole and neutron star X-ray binaries (Maccarone \& Coppi 2003), which is
suggestive of the idea that the AGN pass through similar outburst curves, but
this merely suggestive evidence is the strongest evidence to date that AGN
actually {\it do} have outbursts due to disk instabilities, rather than that
they merely should have such outbursts.

\section{Mass loss from accretion disks}
The evidence for disk winds in active galactic nuclei has been well-reviewed
by Ken Pounds in this volume.  In X-ray binaries, similar types of evidence
for disk winds -- the discovery of X-ray absorption lines which appear to be
dependent on inclination angle and on source spectral state (e.g. Diaz-Trigo
et al. 2006; Neilsen \& Lee 2009; Ponti et al. 2012).  X-ray binaries provide
an additional means of searching for evidence for disk winds.  The mass
transfer rates can be estimated from both the luminosities of the hot spots in
quiescence, and from binary evolution modelling.  These can then be compared
with the long-term mean luminosities of the system, which provide an
independent estimate of the mass accretion rate.  If the mean mass transfer
rate from the donor star is substantially larger than the mean mass accretion
rate by the accretor, then there is additional evidence in support of winds
being important.  This methodology, too, shows that substantial mass loss is
likely to be taking place from accretion disks around black holes
(e.g. Froning et al. 2011).

\section{Gravitationally lensed quasars}

Ideally, one would like to learn about the structures of accretion
disks by imaging them, rather than by testing models of their spectral
or variability properties.  Good prospects exist for making millimeter
VLBI images of a very small number of very nearby galactic nuclei
(e.g. Doeleman 2008).  Relatively little hope exists in the near term
for making direct images of a large sample of black hole accretion
disks spanning a range of flux levels, and relatively little hope
exists for doing small scale direct imaging at frequencies other than
in the millimeter through sub-millimeter band.\footnote{In the longer
  wavelength radio bands, an alternative indirect imaging technique
  has recently been applied -- the examination of the details of
  interstellar scintillation properties of a source (Macquart et
  al. 2013).  This technique is useful only for very compact radio
  bright objects -- i.e. core dominated active galactic nucleus jets
  -- and since this article is concerned with disks, we do not discuss
  the technique except to point out that it exists.}

A technique has been exploited for making indirect imaging measurements of the
accretion flows around active galactic nuclei -- namely using gravitationally
lensed quasars (see e.g. Chen et al. 2012 for an extensive discussion of
recent results; Chang \& Refsdahl 1979 for the first discussion of the
possibility).  When a quasar is behind a galaxy or group or cluster of
galaxies, two types of lensing take place.  The first is that the effect of
the ``smooth mass'' of the lens (i.e. the lens's dark matter halo plus the sum
of the stars) leads to the production of multiple images of the background
object.  The second is that individual stars in the lensing galaxies may
microlens the background object.  For cases where the lensed background source
is much larger than the stars, the effects of the microlensing process are
small, since then only a small part of the object is microlensed at a time.
As a result, the magnitude of variability due to microlensing can be used to
probe the size scale of the background object relative to the sizes of the
stars doing the lensing.

By observing the amplification factors due to microlensing at different
wavelengths, one can map out the size scale of the accretion disks versus
wavelength.  The technique of choice for such work is time series analysis of
microlensing.  In the ideal case, the system will be well enough studied that
the time delays due to the different path lengths light travels to form each
of the observed multiple images are known.  Then one can correct for these
time delays, and remove the variability instrinsic to the quasar, so that the
variability due to microlensing can be isolated (e.g. Morgan et al. 2008,
2010).   

It can be difficult to arrange large numbers of epochs of monitoring data in
the X-rays, particularly when arcsecond angular resolution is needed, and only
Chandra can provide the necessary data.  The optical monitoring data are more
readily obtained.  As a result, methods which invoke less intensive X-ray
coverage are desirable.  Pooley et al. (2006) show that optical data can be
used to determine the magnitude of the optical microlensing anomalies, and
then a single X-ray epoch can be used to estimate the variance in the X-ray
magnifications due to microlensing in a statistical sense, from the variance
in the X-ray images' brightnesses.  This can still often yield important
information about the size scale of the X-ray emitting region, while using
considerably less time on the most oversubscribed telescopes.

A few key results come from Morgan et al. (2008;2010)'s studies: that the
optical continuum comes from regions with spatial scales of $\sim100r_g$, that
the X-ray continuum comes from regions with spatial scales of $\sim10r_g$, and
that the Fe K$\alpha$ emission typically comes from regions even smaller than
the X-ray continuum.  Microlensing of the broad line regions of quasars -- by
the sheer fact that any microlensing is detected at all -- indicates that the
broad line regions are not spherically symmetric (Sluse et al. 2012).

Chen et al. (2013) have shown that the gravitational lensing by the quasar's
own black hole can lead to factor of $\sim2$ systematic errors on the
estimates of the spatial scales .  This result applies primarily to the small
X-ray emission regions, since for much spatial scales of more than tens of
Schwarzschild radii, the effects of light bending by the black hole are very
small.  Usually, the size of the emission region will be under-estimated, but
the direction and magnitude of the effect depend on the inclination angle of
the accretion disk, the spin of the black hole, and the emissivity profile of
the accretion disk.  These errors are of the same order as errors in black
hole masses from most techniques used for active galactic nuclei.  Chen et
al. (2013) also find that subtle differences in the time delays for different
images should be detectable with excellent microlensing campaigns.  With good
enough data, the inclination angles and spins of black holes in quasars might
be measureable using microlensing -- giving measurements independent of those
which come e.g. from iron line measurements.

It is important to note that low luminosity active galactic nuclei have not
yet been well-studied using these techniques, and may have different accretion
geometries -- the systems which have been analyzed are all bright quasars.
The finding of very small X-ray emission regions thus does {\it not} have any
clear implications for the controversy about whether the thin accretion disks
around black holes in X-ray binaries are truncated, with an inner advection
dominated region emitting most of the hard X-rays (see e.g. Rykoff et
al. 2006; Kolehmainen et al. 2013).  These data thus {\it do not} help to
resolve the controversies about whether the inner accretion disks in low/hard
states are truncated, or extend in to the innermost stable circular orbits --
although the technique is, in principle, useful for resolving such
controversies in low luminosity AGN, if lensed LLAGN can be discovered which
are bright enough to perform such studies.  While there has been a recent
discovery of a candidate low luminosity AGN with short time delays between the
different images and flux ratio anomalies (Anguita et al. 2009), this
particular object is extremely faint (the brightest image is seen at magnitude
24.6 in the 606W and 814W filters with the Hubble Space Telescope), and it is
unlikely that it will be useful for understanding the X-ray geometry of low
luminosity AGN.

\section{Summary}
Several key pieces of accretion disk theory show good agreement
between models and observations -- the basic structures of the
accretion disks as geometrically thin, and optically thick in their
outer parts, and the existence of hot spots and spiral arms are all
well established.  At the same time, there is much about accretion
disks which is of great importance which is not fully understood --
particularly the development of a theory of why and how accretion
disks vary.  A great deal of the observed phenomenology agrees with
the basic picture of the hydrogen ionization instability model.
Particularly, the accretion rate above which systems becomes stable is
in good agreement with the predictions of that model, as is the
relationship between outburst peak luminosities and system orbital
periods.  At the same time, there are observational results, such as
the duration of the outbursts of several short period X-ray binaries,
that probably require mass transfer variations as well, and the
detailed shapes of the outbursts are not always well matched by
models.

%
%

\begin{acknowledgements}
The author thanks the conference organizers for having promoted a series of
stimulating discussions.  He also thanks the Avett Brothers,, whose sublime
{\it Magpie and the Dandelion} made the process of finalizing this manuscript
far more enjoyable than it would have been otherwise. 
\end{acknowledgements}



\begin{thebibliography}{}
\bibitem{1}Albright G.E., Richards M.T., 1996, ApJ, 459, L99
\bibitem{104}Anguita T., Faure C., Kneib J.-P., Wambsganss J., Knobel C., Koekemoer A.M., Limousin M., 2009, A\&A, 507, 35
\bibitem{192}Bailey J., Ward M., 1981, MNRAS, 194P, 17
\bibitem{95}Baptista R., Horne K., Hilditch R., Mason K.O., Drew J.E., 1995, ApJ, 448, 395
\bibitem{197}Baptista R., Horne K., Wade R.A., Hubeny I., Long K., Rutten R.G.M., 1998, MNRAS, 298, 1079
\bibitem{96}Baptista R., 2001, LNP, 572, 307
\bibitem{2}Bianchini A., 1990, AJ, 99, 1941
\bibitem{3}Bondi H., Hoyle F., 1944, MNRAS, 104, 273.
\bibitem{4}Cannizzo J.,K., 1998, ApJ, 493, 426
\bibitem{5}Cantrell A.G., et al., 2010, ApJ, 710, 1127
\bibitem{119}Chang K., Refsdahl S., 1979, Nature, 282, 561
\bibitem{6}Chen B., Dai X., Kochanek C.S., Chartas G., Blackburne J.A., Morgan
  C.W., 2012, ApJ, 755, 24
\bibitem{111}Chen B., Dai X., Baron E., Kantowski R., 2013, ApJ, 769, 131
\bibitem{7}Chen W., Shrader C.R., Livio M., 1997, ApJ, 491, 312
\bibitem{102}Church M.J., 2004, RevMexAA, 20, 143
\bibitem{10}Church M.J., Ba\/luci\'nska-Church M., 2004, MNRAS, 348, 955
\bibitem{70}Corbet R.H.D., 1986, MNRAS, 220, 1047
\bibitem{8}Coriat M., Fender R.P., Dubus G., 2012, MNRAS, 424, 1991
\bibitem{9}Crawford J.A., Kraft R.P., 1956, ApJ, 123, 44
\bibitem{11}Curtis H., PLicO 13, 55 (1918).
\bibitem{88}Diaz-Trigo M., Parmar A.N., Boirin L., Mendez M., Kaastra J.S., 2006, A\&A, 445, 179
\bibitem{997}Doeleman S, 2008, JPhCS, 131, 2055
\bibitem{12}Done C., Davis S.W., Jin C., Blaes O., Ward M., 2012, MNRAS, 420, 1848
\bibitem{91}Eracleous M., Halpern J.P., 2003, ApJ, 599, 886
\bibitem{13}Froning C.S., et al., 2011, ApJ, 743, 26
\bibitem{14}Ghosh P., Lamb F.K., 1979, ApJ, 234, 296
\bibitem{15}Godon P., Sion E.M., 2005, MNRAS, 361, 809
\bibitem{16}Greenstein J.L., Kraft R.P., 1959, ApJ, 130, 99
\bibitem{17}Hameury J.-M., King A.R., Lasota J.-P., 1986, A\&A, 162, 71
\bibitem{18}Hameury J.-M., Viallet M., Lasota J.-P., 2009, A\&A, 496, 413
\bibitem{19}Harpaz A., Rappaport S., 1991, ApJ, 383, 739
\bibitem{141}Haswell C.A., King A.R., Murray J.R., Charles P.A., 2001, MNRAS,
  321, 475
\bibitem{92}Hemphill P.B., Rothschild R.E., Caballero I., Pottschmidt K., K\"uhnel M.,F\"urst F., Wilms J., 2013, ApJ, 777, 61
\bibitem{20}Hjellming R. et al., 2000, ApJ, 544, 977
\bibitem{97}Horne K., 1993, in Accretion Disks in Compact Stellar Systems, ed J.C. Wheeler (Singapore:World Scientific), 117
\bibitem{21}Hut P., Paczynski B., 1984, ApJ, 284, 675
\bibitem{500}Hynes R.I., et al., 2013, ApJ, in press
\bibitem{998}Jonker P., van der Klis M., 2001, ApJ, 553L, 43
\bibitem{22}Joy A.H., 1954, ApJ, 120, 377
\bibitem{23}Joy A.H., 1956, ApJ, 124, 317
\bibitem{24}K\"ording E.G., Jester S., Fender R., 2006, MNRAS, 372, 1366
\bibitem{25}King A.R., Ritter H., 1998, MNRAS, 293, 42
\bibitem{800}Knevitt G., Wynn G., Vaughan S., Watson M., 2013, MNRAS, in press
\bibitem{101}Knigge C., Long K. S.,Wade R. A., Baptista R., Horne K., Hubeny
  I., Rutten R. G. M., 1998a, ApJ, 499, 414
\bibitem{122}Kolehmainen M., Done C., Diaz Trigo M., 2013, MNRAS, in press.
\bibitem{26}Kuiper G.P., ApJ, 93, 133, (1941).
\bibitem{27}Kraft R.P., 1964, ApJ, 139, 4578
\bibitem{503}La Barbera A., Burderi L., Di Salvo T., Iaria R., Robba N., 2001,
  ApJ, 553, 375
\bibitem{28}Lasota J.-P., 2001, NewAR, 45, 449
\bibitem{29}Maccarone T.J., Coppi P.S., 2003, MNRAS, 338, 189
\bibitem{82}Maccarone T.J., Kundu A., Zepf S.E., Rhode K.L., 2007, Nature, 445, 183
\bibitem{30}Maccarone T.J., Long K.S., Knigge C., Dieball A., Zurek D.R., 2010, MNRAS, 406, 2087
\bibitem{181}Maccarone T.J., Patruno A., 2013, MNRAS, 428, 1335
\bibitem{400}Macquart J.-P., Godfrey L.E.H., Bignall H.E., Hodgson J.A., 2013,
  ApJ, 2013, ApJ, 765, 142
\bibitem{170}Maitra D., Bailyn C.D., 2008, ApJ, 688, 537
\bibitem{94} Marsh T.R., 1988, MNRAS, 231, 1117
\bibitem{31}McClintock J.E., Horne K., Remillard R.A., 1995, ApJ, 442, 358
\bibitem{32}Miller-Jones J.C.A., Jonker P.G., Dhawan V., Brisken W., Rupen M.P., Nelemans G., Gallo E., 2009, ApJ, 706L, 230
\bibitem{33}Miller-Jones J.C.A., Sivakoff G.R., Knigge C., K\"ording E., Templeton M., Waagen E.O., 2013, Science, 340, 950
\bibitem{34}Mitsuda K., et al., 1984, PASJ, 36, 741
\bibitem{35}Morgan C.W., Kochanek C.S., Dai X., Morgan N.D., Falco E.E., 2008, ApJ, 689, 755
\bibitem{36}Morgan C.W., Kochanek C.S., Morgan N.D., Falco E.E., 2010, ApJ, 712, 1129
\bibitem{37}Neilsen J., Lee J.C., 2009, Nature, 458, 481
\bibitem{199}O'Donoghue D., Charles P.A., 1996, MNRAS, 282, 191
\bibitem{87}Olson E.C., 1991, AJ, 102, 1423
\bibitem{38}Orosz J.A., et al., 2001, ApJ, 555, 489
\bibitem{110}Orosz J.A., et al., 2007, Nature, 449, 872
\bibitem{39}Osaki Y., 1996, PASP, 108, 39
\bibitem{160}Osaki Y., Kato T., 2013, PASJ, in press
\bibitem{40}Pandel D., Cordova F.A., Howell S.B., 2003, MNRAS, 346, 1231
\bibitem{41}Patterson J., 2011, MNRAS, 411, 2695
\bibitem{103} Pietsch W., et al., 2006, ApJ, 646, 420
\bibitem{42}Piraino S., Santangelo A., Ford E.C., Kaaret P., 1999, A\&A, 349L, 77
\bibitem{43}Pooley D., Blackburne J.A., Rappaport S., Schecheter P.L., Fong W-f., 2006, ApJ, 648, 67
\bibitem{44}Ponti G., Terrier R., Goldwurm A., Belanger G., Trap G., 2010, ApJ, 714, 732
\bibitem{45}Ponti G., Fender R.P., Begelman M.C., Dunn R.J.H., Neilsen J., Coriat M., 2012, MNRAS, 422, L11
\bibitem{46}Portegies Zwart S.F., Dewi J., Maccarone T., 2004, MNRAS, 355, 413
\bibitem{47}Reid M.J., McClintock J.E., Narayan R., Gou L., Remillard R.A., Orosz J.A., 2011, ApJ, 742, 83
\bibitem{294}Reimers D., Hagen H.-J., 2000, A\&A, 358L, 45
\bibitem{48}Robinson E.L., Wood J.H., Wade R.A., 1999, ApJ, 514, 952
\bibitem{394}Rutten R.G.M.R., van Paradijs J., Tinbergen J., 1992, A\&A, 260, 213
\bibitem{121}Rykoff E.S., Miller J.M., Steeghs D., Torres M.A.P., 2007, ApJ,
  666, 1129 
\bibitem{49}Schatzman E., 1949, AnAp, 12, 281
\bibitem{50}Schreiber M.R., Hemury J.-M., Lasota J.-p., 2003, A\&A, 410, 239
\bibitem{52}Seyfert C.K., ApJ, 97, 28 (1943).
\bibitem{51}Shahbaz T., Charles P.A., King A.R., 1998, MNRAS, 301, 382
\bibitem{53}Shakura N.I., Sunyaev R.A., 1973, A\&A, 24, 337
\bibitem{931}Sluse D., Hutsemekers D., Courbin F., Meylan G., Wambsganss J.,
  2012, A\&A, 544A, 62
\bibitem{54}Smak J.I., 2000, AcA, 50, 399
\bibitem{71}Sokoloski J., 2003, JAAVSO, 31, 89
\bibitem{90}Soria R., 2002,  Proceedings of the MGIX MM Meeting, eds. Robert T Jantzen \& Remo Ruffini, World Scientific Publishing Company
\bibitem{55}Steeghs D., 23001, LNP, 573, 45
\bibitem{85}Thomas H.-C., 1977, ARA\&A, 15, 127 
\bibitem{300}Truss M.R., Wynn G.A., Murray J.A., King A.R., 2002, MNRAS, 337, 1329
\bibitem{151}Wachter S., Hoard D.W., Bailyn C.D., Corbel S., Kaaret P., 2002,
  ApJ, 568, 901
\bibitem{56}Walker M.F., 1954, PASP, 66, 230
\bibitem{57}Walker M.F., 1956, ApJ, 123, 68
\bibitem{145}Wang Z., Chakrabarty D., 2010, ApJ, 712, 653
\bibitem{501}Warner B., 1983, ASSL, 101, 155
\bibitem{58}Warner B., 1987, MNRAS, 227, 23
\bibitem{59}White N.E., Mason K.O., 1985, SSRv, 40, 167
\bibitem{60}White N.E., Peacock A., Hasinger G., Mason K.O., Manzo G., Taylor B.G, Branduardi-Raymont G., 1986, MNRAS, 218, 129
\bibitem{131}Whitehurst R., 1988, MNRAS, 232, 35
\bibitem{502}Wijnands R., van der Klis M., 1998, Nature, 394, 344 
\bibitem{105}Wijnands R., Miller J.M., Markwardt C., Lewin W.H.G., van der Klis M., 2001, ApJ, 560L, 159
\bibitem{93}Wood J.H., Abbott T.M.C., Shafter A.W., 1992a, ApJ, 393, 729
\bibitem{100}Wood J.H., Horne K., Vennes S., 1992b, ApJ, 385, 294
\bibitem{61}Wood J.H., Horne K., Berriman G., Wade R.A., 1989, 341, 974
\bibitem{99}Wood J., Horne K., Berriman G., Wade R., O'Donoghue D., Warner B., 1986, 219, 629
\bibitem{62}Wu Y.X., Yu W., Li T.P., Maccarone T.J., Li X.D., 2010, ApJ, 718, 620
\bibitem{63}Zdziarski A.A., Wen L., Gierlinski M., 2007, MNRAS, 377, 1006
\bibitem{80}Zepf S.E., Stern D., Maccarone T.J., Kundu A., Kamionkowski M., Rhode K.L., Salzer J.J., Ciardullo R., Gronwall C., 2008, ApJL, 683, 139
\bibitem{171}Zurita C., et al., 2002, MNRAS, 333, 791
\bibitem{64}Zurita C., Durant M., Torres M.A.P., Shahbaz T., Casares J., Steeghs D., 2008, ApJ, 681, 1458
\end{thebibliography}
\end{document}